\documentstyle[12pt]{article}
\def\be{\begin{equation}}
\def\ee{\end{equation}}
\def\({\left (}
\def\){\right )}
\def\[{\left [}
\def\]{\right ]}

\def\tl{\tilde{\lambda}}

\begin{document}
\renewcommand{\thefootnote}{\fnsymbol{footnote}}
 \begin{titlepage}
\vspace*{-4ex}
 February, 1993 \hfill LBL--33833\\
 \null \hfill hep-ph/9303282 \\

\vskip 1 true cm
 \begin{center}
 {\bf \LARGE  First Order Phase Transition in a Two \\[0.1cm] Higgs-doublet
Model with $M_{h}>M_{W}$\footnote{This work was supported by the
Director, Office of Energy Research, Office of
High Energy and Nuclear Physics, Division of High Energy Physics
of the U.S. Department of Energy under contract DE-AC03-76SF00098 and in part
by the National Science Foundation under grant PHY--90--21139.}}\\[7ex]
{\large
  Vidyut Jain and Aris Papadopoulos}
   \\ [2ex]
%

%$^1$
{\large \it  Theoretical Physics Group,\\
 Lawrence Berkeley Laboratory,\\ 1 Cyclotron Road,
  Berkeley, California 94720}\\ [2ex]
 \end{center}
\vskip 1.4cm
 \begin{abstract}
 We study the 3d effective field theory of a weakly coupled
 two Higgs-doublet
 scalar model at high temperature. Our model has
three scalar quartic couplings and an $O(4)$
symmetry which is
 spontaneously broken by a nonzero scalar field $vev$ at zero
temperature. Using high temperature perturbation theory,
 renormalization group arguments in $4-\epsilon$
dimensions, as well as $1/N$ expansion techniques in 3 dimensions,
we argue that the  transition from the high temperature
symmetry restoring phase to the
 low temperature phase is first order for a range of scalar
couplings. This result is not due to gauge couplings.
We discuss the implications of our results for the
transition
in the two Higgs-doublet electroweak model, especially when the
lightest neutral Higgs is heavier than the W--bosons.
 \end{abstract}
%

%\end{titlepage}
%\end{document}
\newpage
%%%%%%%%%%%%%%%%%%%%%%%%%%%%%%%%%%%%%%%%%%%%%%%%%%%%%%%%%%%%%%%
\renewcommand{\thepage}{\roman{page}}
\setcounter{page}{2}
\mbox{ }
\vskip 1in
\begin{center}
{\bf Disclaimer}
\end{center}
\vskip .2in
\begin{scriptsize}
\begin{quotation}
This document was prepared as an account of work sponsored by the United
States Government. Neither the United States Government nor any agency
thereof, nor The Regents of the University of California, nor any
of their employees, makes any warranty, express or implied, or assumes
any legal liability or responsibility for the accuracy, completeness, or
usefulness of any information, apparatus, product, or process disclosed,
or represents that its use would not infringe privately owned rights.
Reference herein to any specific commercial products process, or service by its
trade name, trademark, manufacturer, or otherwise, does not necessarily
constitute or imply its endorsement, recommendation,or favoring by the
United States Government or any agency thereof, or The Regents of the
University of California. The views and opinions of authors expressed herein
do not  necessarily state or reflect those of the United States Government
or any agency thereof of The Regents of the University of California and
shall not be used for advertising or product endorsement purposes.

\end{quotation}
\end{scriptsize}
\vskip 2in
\begin{center}
\begin{small}
{\it Lawrence Berkeley Laboratory is an equal opportunity employer.}
\end{small}
\end{center}
\newpage

 \end{titlepage}
\setcounter{page}{1}

{\bf Introduction}.
The topic of high temperature phase transitions in 4d
scalar models has attracted a lot of interest recently due to
the possibility of baryogenesis during the weak phase transition.
It has been argued that baryogenesis requires a sufficiently strong
first order phase transition [1,2,3,4].

The nature of the phase transition can be determined if one has
a reliable determination of the finite temperature effective
scalar potential $V_{eff}$.
\footnote{What we mean by $V_{eff}$ is the
sum of one-particle irreducible diagrams with no external legs. This
quantity need be neither real nor convex [5], in contrast to the
standard definition of the effective potential given in many textbooks [6].
However, the minimum of the real part of our effective potential still
gives the ground state of the theory.}
It has been known for some time [7,8,9] that at sufficiently
high temperature the ground state of a $\phi^4$
 model is symmetry preserving with $<\phi>=0$,
even if the zero
temperature ground state is not. What is not well understood is
how the transition from the high temperature phase to the low temperature
phase proceeds in generic models with several couplings.

One can imagine two possibilities as the temperature is lowered from  a very
high value: i) the minimum remains at $<\phi>=0$ until a temperature
$T_1$ at which point it jumps discontinuously to nonzero values
$|<\phi>|\neq 0$, or ii) the minimum remains at $<\phi>=0$ until a
temperature $T_2$ at which it moves continuously away from the origin
as the temperature is lowered further. Case i) indicates a first order
transition and case ii) indicates a second order transition.

The  determination of the transition order
is not always easy due to infrared divergences in high $T$
field theory which can render perturbation theory useless in calculating
$V_{eff}$ too close to the origin and at temperatures near
the transition temperature. In this case, one can i) try resuming
perturbation theory or ii) use
renormalization group (RG) methods or iii)
use $1/N$ expansion techniques to
gain some insight. When a model has several couplings, any one method
cannot be trusted to give a reliable description of the transition for all
 values  of the (perturbative) couplings.

The most thoroughly investigated electroweak model, the minimal
standard model, can be argued [10-14] to give a first
order transition for $M_{Higgs}$ sufficiently
smaller than $M_W$. In this case, high $T$
perturbation theory can give a good description of the transition if
used carefully.

The leading expression for the high temperature
effective potential when $M_{Higgs}\ll M_{W}$ is of the form [7,15,10--14]
\be  V_{eff} = m^2(T)|\Phi|^2- \delta T |\Phi^3| +
   {\lambda(T)\over 4!}|\Phi|^4,
\ee
where $\Phi$ is the Higgs doublet, $\delta$ is proportional to the
cube of the $SU(2)$ gauge coupling and $m^2(T)\propto (T^2-T_2^2)$.
This potential describes a first order transition whose minimum jumps
discontinuously to a value $\propto \delta T/\lambda$ at a temperature
$T_1>T_2$.
RG methods combined with $\epsilon$--expansion techniques [15,16] as well as
computations of $V_{eff}$ using
$1/N$ techniques (when $M_{W}\sim N M_{Higgs}$)
[17,18] are consistent with this result.

With the current experimental bounds on $M_{Higgs}$ [19], the
situation $M_{Higgs}$ $\geq O(M_W)$ is the one of physical interest.
Unfortunately,
for $M_{Higgs} \geq M_W$, perturbation theory breaks down and
 there is really no clear picture of what happens.
However,
one expects that at best the phase transition will be very weakly first order.

It has been suggested [11] that one should augment the minimal standard model
with extra scalars in order to produce a big enough first order effect for
values of the Higgs mass larger than the current experimental bound.
The idea is to enlarge the ratio $\delta/\lambda$ during the phase transition.

One suggestion [11], is to add an electroweak singlet scalar $S$
with a large enough
coupling to the Higgs doublet $\Phi$ of the form $\xi \Phi^{\dag}
\Phi S^2$.
It has been argued that if $S$ is massive enough it
contributes to $V_{eff}(T=0)$ in a way that reduces
the value of $\lambda$ without decreasing $M_{Higgs}$.

Another nonminimal model is the two Higgs--doublet electroweak model.
Due to the number of unknown parameters in this model it makes
sense to study  a toy model before attempting to understand
the nature of the transition in the general case. In this
letter we will study such a toy model. In particular,
we study the 4d weakly coupled scalar model with
tree potential
\def\psib{\phi_2}\def\phib{\phi_1}
\be V(\phib,\psib)={\lambda\over 4!} \[ (\phib^2-v^2)^2+(\psib^2-v^2)^2\]
 +{\xi\over 12} \phib^2 \psib^2 + {\alpha\over 6}
 (\phib\cdot \psib)^2, \ee
where $\lambda>\xi$ (to ensure real physical masses) and
$\phib$ and $\psib$ are both $O(4)$ vectors. Unless otherwise stated,
we assume all couplings are arbitrarily small (their ratios, however,
may not be) and nonnegative.

The model (2) has an $O(4)$ symmetry as well as the discrete symmetry
$\phib \leftrightarrow \psib$. It is also invariant under $\phib \rightarrow
-\phib$ $or$ $\psib\rightarrow-\psib$. The potential (2) is a subset of the
general potential describing the two Higgs-doublet
model which has five quartic couplings [20], not three. One of
the additional couplings breaks the $Z_2$ symmetry $\phib\leftrightarrow
\psib$, the other is allowed because the potential needs to be invariant
only under $SU(2)\times U(1)$, not $O(4)\sim SU(2)\times SU(2)$.

Although the 2 Higgs-doublet model has been studied at one--loop by previous
authors [21], there has never been a reliable demonstration of a first order
transition when the lightest neutral Higgs weighs at least  $O(M_W)$.

In this article we argue that perturbation theory gives a reliable
description of a first order transition if $\alpha/\lambda$ and
$\alpha/\xi$ are sufficiently big. We present additional arguments for
a first order transition based on RG group methods
combined with the $\epsilon$--expansion, as well as computation of $V_{eff}$
based on $1/N$ expansion techniques.
We stress that
the first order behavior is not due to gauge fields. We will later
address the implications of our results for baryogenesis in the two
Higgs--doublet electroweak model.

We study not the full 4d model, but the 3d effective field theory
describing the high $T$ limit of the 4d model. Before proceeding with
our analysis we introduce all the tools we will use in the context
of a simple  $\phi^4$ model.

{\bf Effective 3d Theory. }
Let us consider the ungauged scalar theory with 4d potential
\be V(\phi) ={\lambda\over 4!} (\phi^2-v^2)^2. \ee
The model is, at high $T$, formally
equivalent to a euclidean field theory with one compact dimension. For
physics at scales less than $O(2\pi T)$ it is sufficient to study the
effective 3d Lagrangian that results from integrating out the compact
dimension. The 4d fields can be expanded as
\be \phi(\vec{x},\tau)=\sum_n \phi_n(\vec{x})\psi^n(\tau), \ee
where $\tau$ parameterizes the compact dimension and the $\psi^n(0)
=\psi^n(T^{-1})$. Thus the 4d field yields an infinite tower of 3d
fields when the compact dimension is integrated out in the action.
Furthermore, for what we are interested in, only the zero model $n=0$,
for which $\psi^0(\tau)=\psi^0(0)$ is important in the effective 3d model
because the others pick up large $O(2\pi T)$ masses from the 4d kinetic term.

The largest effect of integrating the nonzero modes out from the theory
is to give an $O(\lambda T^2)$ correction to the mass--squared of the
zero mode [16]. Therefore, we study an effective 3d theory with potential
\def\vp{\varphi}\def\tl{\tilde{\lambda}}\def\tv{\tilde{v}}
\be V(\vp) = {\tl\over 48}(T-4\tv^2)\vp^2+{\tl\over 4!}\vp^4 , \ee
where we have defined the 3d quantities
\be \vp=\phi_0/\sqrt{T},\qquad \tl=\lambda T,\qquad \tv^2= v^2/T.\ee

The effective theory can be studied by computing the 3d $V_{eff}$, or by
RG methods.
The ungauged model is believed to have a
second order transition [22]. This result is not accessible by means of
high temperature perturbation theory. To see this we need only compute
$V_{eff}$ to one--loop.

The field dependent mass for the potential (5) is
\be m^2(\vp,T)= {\tl\over 24} (12\vp^2+T-4\tv^2).  \ee
We expect the 3d model to give a good description of
$V_{eff}(\vp,T)$ for $m^2<4\pi^2 T^2$, i.e.
for $T^2$ near $4v^2$ and
${1\over 2}\tl \vp^2< 4\pi^2 T^2$.
\footnote{We will freely switch back and forth from
3d quantities and 4d quantities, dropping the zero mode subscripts.}

The one--loop contribution to the effective potential is then ($m^2\geq 0$)
\be {1\over2} {\rm Tr}\ln [-\vec{\partial}^2+m^2] =
    {\Lambda m^2\over 4\pi^2} - {|m|^3\over 12\pi} + {\rm const.}
\ee
We have used a sharp momentum cutoff to regulate the integral.

The linearly divergent field dependent term is absorbed by renormalizing
the 3d theory. The field dependent part of the one--loop effective potential
is then
\be V_{eff}= {\tl\over 48}(T-4\tv^2)\vp^2-{|m|^3\over 12\pi}
  +{\tl\over 4!}\vp^4 .\ee
This result reproduces the leading temperature dependent part of the
so--called daisy sum in the high temperature field theory. Taken at
face value, (9) indicates a first order transition with a discontinuous
jump in the minimum $|<\phi>| \propto \sqrt{\lambda} T$ at a temperature
$T_1$ slightly above $T_2=2v$. Since we have already indicated that
the phase transition is believed to be second order, eq. (9) must not
give a reliable determination of $V_{eff}$. In fact, it is well known
[8,9] that higher loop corrections are significant.
In 3d, the scalar coupling
$\tl$ has dimension 1. The dimensionless perturbation parameter is
actually $\sim\tl/m$, a result that can be checked by examining multiloop
graphs [8,9]. At $T=T_2$, $\tl/m\sim \sqrt{\lambda}T/\phi$, and for the
apparent minimum at $|<\phi>|\propto \sqrt{\lambda} T$ this
expansion parameter is of $O(1)$ and therefore perturbation theory cannot
be relied upon to distinguish between  1st and 2nd order behaviour.

In the case of an $O(N)$ symmetric model described by (3), $V_{eff}$
has been computed using $1/N$ expansion techniques [23,24,22,8] to
next-to-leading order.
The results indicate that the phase transition is second order
[24,17]. This method quickly
becomes technically complicated, so that $V_{eff}$ has not been computed
beyond next-to-leading order (known
results should not be trusted to give a reliable description
of what happens for low values of $N$).

Fortunately, there have been many studies of scalar models based on the
renormalization group method [22].
A trajectory in coupling constant
space which leads to an infrared stable fixed point under the RG
is identified with 2nd order behaviour. There
are two possibilities to study the RG flow equations for a
3d theory.
One is to work in 4-$\epsilon$ dimensions to some order in $\epsilon$.
Although one is interested in $\epsilon=1$, even lowest order results
in $\epsilon$ can give useful insight into the nature of the phase
transition.\footnote{ The
absence of an infrared stable fixed point
at $O(\epsilon)$ although not
a proof [25] that the model has first order behaviour,
may hint at such behaviour for some range of couplings.}
The other is to stay in 3d and to work to some order in $1/N$. As with $1/N$
computations of $V_{eff}$, leading order results in this expansion are
generally only expected to give a good description of the model when there
are many scalars.

For a simple
$O(N)$ symmetric scalar theory, RG equations have been studied to
a high order in both the $1/N$ expansion and the $\epsilon$--expansion [22].
Both results indicate a 2nd order transition and, in both cases,
leading order results give a faithful indication of this.
For example, for $N=1$ and 4d potential (3), the
$O(\epsilon)$ and $O(\lambda^2)$ RG flow equation is [16,22,26]
\be
   {d\lambda\over dt} = \epsilon\lambda-{3\over 16\pi^2}\lambda^2.
\ee
Here, $t$ increases in the infrared. This has an infrared unstable
fixed point at $\lambda^*=0$ and an infrared stable fixed point at
$\lambda^*=16\pi^2\epsilon/3$. In the infrared limit, every positive
$\lambda$ is driven to this point.

For an abelian Higgs model, i.e. a gauged $O(2)$ model, with gauge
coupling $g$ and 4d potential (3) where $\phi^2=\phi\cdot\phi$, the
RG equations to $O(\epsilon)$
have no infrared stable fixed point [15,16], suggesting that
a first order transition may be possible. In fact, high $T$
perturbation theory can be used to give a reliable description of a
1st order transition for $g^2\gg \lambda \gg g^4$ [15,7]. Perturbation
theory is reliable because the perturbative
expansion parameters  go like some (positive) power of $\lambda/ g^2$
at both minima at $T=T_1$ [10,12,14].

For $M_{Higgs}\geq M_W$ the above analysis breaks down. For the abelian
Higgs model, a computation of $V_{eff}$ to next-to-leading order [17,18]
in the $1/N$ expansion gives a second order transition. Futhermore,
calculations of critical exponents during a 2nd order transition
to next--to--leading order in $1/N$ remain physical for values of $N\geq 10$
[29,15] suggesting that the prediction of a second order transition for
$M_W \leq M_{Higgs}$ and $N\geq 10$ is reliable. In fact, the phase transition
is known to be second order for $N=2$, $M_W \leq M_{Higgs}$ due to other
reasons [30]. This demonstrates that the RG analysis to $O(\epsilon)$
in $4-\epsilon$ dimensions can fail to indicate 2nd order behaviour.

As an aside,
we note that much less is well understood when the gauge group
is nonabelian. As with the abelian case, there is no infrared stable
fixed point to $O(\epsilon)$ in $4-\epsilon\; $d. Assuming that infrared
divergences in loops due to the self interactions of the magnetic fields
are cutoff by a nonperturbatively generated $T$-dependent mass [27], it is
again possible to argue [10,12,13] that for $g^2\gg \lambda$ the transition
is first order. For $g^2 <  O(\lambda)$, $1/N$ studies [18,31] suggest that
the transition may be second order for a model with enough scalars.
For a model with just a few scalars, recent results [32] due to resuming
high $T$ perturbation theory in the minimal electroweak model
also suggest 2nd order behaviour for $g^2 <  O(\lambda)$. In addition, it
has been argued [33] that the existence of a fixed point at $O(\epsilon)$
in $2+\epsilon$ d may give a reliable indication of a second order transition
at $d=3$ and low $N$.

{\bf The Toy Model.}
We are finally ready to demonstrate that the
ungauged model described by (2) has a 1st order phase transition for a range
of couplings. To our knowledge, what we present below is new.

To begin, we examine the RG equations for (2) in $4-\epsilon$ dimensions.
To $O(\epsilon)$, one has [26]
\def\psi{\phi_2}
\begin{eqnarray}
  {d\lambda\over dt} &=& \epsilon \lambda -
     {1\over 12\pi^2} \{ 3\lambda^2+\xi^2+\xi\alpha+\alpha^2\}, \nonumber \\
  {d\xi\over dt} &=& \epsilon\xi - {1\over 12\pi^2} \{
    3\lambda\xi +\lambda\alpha+{1\over 2}\xi^2+{1\over2} \alpha^2\},
      \nonumber \\
  {d\alpha \over dt} &=& \epsilon\alpha-{1\over 12\pi^2} \{
    \lambda\alpha+\xi\alpha+{3\over 2}\alpha^2\}.
\end{eqnarray}
This has the following fixed points:
\begin{eqnarray}
   i) & & \alpha^*=0, \xi^*=0, \lambda^*=0, \nonumber \\
  ii) & & \alpha^*=0, \xi^*=0, \lambda^*=4\pi^2\epsilon, \nonumber \\
 iii) & & \alpha^*=0, \xi^*={24\over 13}\pi^2\epsilon,
          \lambda^*={48\over 13}\pi^2 \epsilon.
\end{eqnarray}

Linear stability analysis reveals the following facts. The fixed point $i)$,
the Gaussian fixed point, is unstable in all directions. The fixed points
$ii)$ and $iii)$ are both unstable in the $\alpha$ direction:  in
the infrared limit, small (positive) values of $\alpha$ are driven larger.
The lack of a stable fixed point suggests that the purely scalar model
may undergo a first order transition. Indeed, the fact that the coupling
$\alpha$ is responsible for the lack of stable fixed points suggests that
any first order behaviour will be due to $\alpha$.

We therefore examine high $T$ perturbation theory for $\alpha\gg \lambda$
and $\alpha\gg \xi$. To simplify the analysis,  we consider
\be \lambda = f^3, \qquad \xi = f^3,  \qquad \alpha=f^2, \ee
for small positive $f$.\footnote{ Our
particular assignments are only for numerical convenience and we will
later make clearer for what values of couplings we predict first
order behaviour.} This choice
of couplings also ensures [14] that all mass scales of interest are much less
than $T$, thus allowing us to work with an effective 3d theory. Under
such circumstances, we may now argue that perturbation theory gives a
reliable description of a first order transition. The argument follows
closely the one for the abelian Higgs model.

The effective 3d potential that we study is
\def\vt{\varphi_2}\def\vp{\varphi_1}\def\tf{\tilde{f}}
\be V_{tree}(\vp,\vt)=
    {(f^3 T)\over 4!} (\vp^2+\vt^2)^2
   +{f^2 T\over 6} (\vp\cdot \vt)^2  + {1\over2}(\vp^2+\vt^2) M^2,\ee
where
\begin{eqnarray}
   M^2 &=& {\alpha+5\lambda\over 36} T^2-{\lambda\over 6} v^2 =
        {f^2\over 36} (1+O(f)) (T^2-T_2^2), \\
   T^2_2 &=& 6\lambda v^2/(\alpha+5\lambda) = 6f v^2 (1-O(f)),
\end{eqnarray}
and $\vp,\vt$ are the (properly normalized) zero modes of the
4d fields $\phib,\psi$.

As with the simple scalar model discussed previously, we expect
the effective 3d theory to give a reliable description of the phase
transition as long as all mass scales are less than $2\pi  T$.

The field dependent mass matrix has the eigenvalues
\be f^2\rho_+ + O(f^3),\quad  f^2\rho_-+O(f^3),\quad
     f^2\sigma_+ +O(f^3),\quad  f^2\sigma_- +O(f^3), \ee
where
\begin{eqnarray}
  & & \rho_\pm ={1\over 36}(T^2-T_2^2) \pm {1\over 3}\phib\cdot\psi,
    \nonumber \\
  & & \sigma_\pm   ={1\over 36}(T^2-T_2^2) + {1\over 6} \[
        \phib^2+\psi^2 \pm \sqrt{(\phib^2+\psi^2)^2+12(\phib\cdot\psi)^2}\] .
\end{eqnarray}
The first two eigenvalues each have multiplicity 3.

The one--loop contribution includes a linearly divergent correction which
renormalizes the 3d theory. The remaining finite contribution is
\be V_{1-loop} = -{ f^3\over 12\pi} \( 3\rho_+^{3\over2}
   +3\rho_-^{3\over2}+
     \sigma_+^{3\over2}+\sigma_-^{3\over2} \) +O(f^4). \ee
At one--loop, we have $V_{eff}=V_{tree}+V_{1-loop}$. At $T=T_2$, the
real part of the potential has a minimum at
$   |<\phib>|=|<\psi>|\sim {T\over 15}$
and $\qquad <\phib\cdot\psi>=O(f^2)$.

Taken at face value, to lowest nontrivial order in $f$, $V_{eff}$
implies  that the minimum is away from the origin by the time the mass
at the origin vanishes and hence that a first order transition has
occurred. The dimensionless loop expansion parameters are proportional to
\be {\lambda T\over m}, {\xi T\over m}\sim {f^3 T\over m} \quad
  {\rm and}\quad {\alpha T\over m}\sim {f^2 T\over m} .
\ee
Here $m$ is a field dependent mass eigenvalue. At $T=T_2$, all mass
eigenvalues are zero at the origin so that $V_{eff}$ is not to be
trusted near the origin. To verify that (23) gives a reliable prediction
of a first order transition we
must first demonstrate: i) there is another minimum degenerate
with the one at the origin at some $T=T_1> T_2$, and ii) the loop
expansion parameters are small at both minima at $T=T_1$.

Since the $\phib\cdot\psi$ terms are $O(f^2)$ at the minimum at $T=T_2$,
they do not affect the lowest order (in $f$) minimization conditions
for $<\phib>$ and $<\psi>$. It turns out that for the purposes of
discussing the leading order (in $f$)  characteristics of the phase
transition one is allowed to just ignore these terms.
To lowest order in $f$, the 3d effective potential we study for
$T$ close to $T_2$ is therefore
\be V_{eff}={(f^3 T)\over 4!}(\vp^2+\vt^2)^2+{M^2\over 2} (\vp^2+\vt^2)
  -{ f^3 \over 12\pi} \[{1\over 36} (T^2-T_2^2)+{1\over3}(\phib^2+\psi^2)
     \]^{3\over2}.\ee
Although this contains a cubic term only at $T=T_2$, it still leads to
first order behaviour because for $T$ close enough to $T_2$ the last term
behaves like a cubic term $\propto (\phib^2+\psi^2)^{3\over2}$.
One can verify that
$T_1=T_2+O(f)T_2$ and $|<\phib>|=|<\psi>|=O(T)$ at the second minimum
at $T=T_1$. At the origin, all mass eigenvalues go as $f^3 T_2^2$ and thus the
loop expansion parameters go as $f\sqrt{f}$ and $\sqrt{f}$. Away from
the origin, in the $\phib\cdot\psi=0$ direction, the mass eigenvalues
only increase and the expansion parameters decrease. Therefore, for
sufficiently small $f$, the perturbative expansion is well controlled at
both minima at $T=T_1$. Just below $T=T_1$ the global minimum is away from the
origin; this is the signal of
first order behaviour.

We end our analysis of the model (2) by discussing
three issues:

1) There are values of $\phib\cdot\psi$ for which
some of the field dependent mass eigenvalues of (14) vanish and hence
perturbation theory cannot be reliably used to
compute $V_{eff}(\phib^2,\psi^2,\phib\cdot\psi)$ for such values.
For example at $T=T_2$, six of the $O(f^2)$ mass eigenvalues
(17),(18) vanish at $\phib\cdot\psi=0$ suggesting that perturbation theory
is not reliable at the apparent minimum away from the origin. Actually,
at $T=T_2$ and $\phib\cdot\psi=0$ these mass eigenvalues go like
$O(f^3 T_2^2)$ so that perturbation theory is reliable at the
minimum. However, there are still values $\phib\cdot\psi=O(f\phib^2,f\psi^2)$
for which the mass eigenvalues vanish and perturbation theory must be
resummed in order to get a more reliable $V_{eff}$.

2) So far we have said nothing about the transition for $\lambda\sim
\xi\sim\alpha$. In this case, since perturbation theory breaks down, we
must resort to other methods. Even though the
RG analysis in $4-\epsilon$ d has no infrared stable fixed point to
$O(\epsilon)$ the transition may be second order for these values of
the couplings. To learn more we can use $1/N$ expansion techniques [22,23]
when $\phib,\psi$ are $N$-vectors.
We have studied two interesting limits:
\begin{eqnarray}
  i) & & \xi=\lambda \;{\rm and}\;  (N^2\lambda), (N\alpha)
     \; {\rm fixed \; for\; increasing} \; N,\nonumber \\
  ii) & & \alpha=\xi=\lambda \; {\rm and}\; (N\lambda)
  \; {\rm fixed\; for\; increasing}\; N.
\end{eqnarray}
At large $N$, case $(22i)$ leads to first order behaviour while
case $(22ii)$
leads to second order behaviour.

To show this last statement it is convenient [23,17,18]
to perform the rescalings
$ \phib\rightarrow \sqrt{N}\phib,\psi\rightarrow\sqrt{N}\psi$
and $v\rightarrow
\sqrt{N}v$. Then, the 3d effective theory is described
by the tree potentials
\begin{eqnarray}
 i) & & V= \[ {T^2\over 72}(\lambda+\alpha)
      -{\lambda\over 12} v^2 \] (\vp^2+\vt^2) + {\lambda T\over 4!}
(\vp^2+\vt^2)^2+{\alpha N T\over 6}(\vp\cdot\vt)^2, \nonumber \\
 ii) & & V= N\[ {T^2\over 72}\lambda
      -{\lambda\over 12} v^2 \] (\vp^2+\vt^2) + {\lambda NT\over 4!}
(\vp^2+\vt^2)^2+{\lambda N T\over 6}(\vp\cdot\vt)^2 \nonumber \\ \;
& & \qquad +O(1) \end{eqnarray}
for $\lambda\rightarrow \lambda/ N^2,\alpha\rightarrow\alpha/N$
and $\lambda\rightarrow \lambda/N$, respectively.
The 3d kinetic terms for both cases are also proportional to $N$, due
to the rescalings of the fields.

To compute $V_{eff}$ in the $1/N$ expansion one performs the shift
\be \vp\rightarrow \vp+\hat{\vp}/\sqrt{N}, \qquad
     \vt\rightarrow \vt+\hat{\vt}/\sqrt{N}, \ee
dropping all terms linear in the quantum fields $\hat{\phib},\hat{\psi}$
and computes consistently
all one-particle-irreducible diagrams with zero external quantum
fields to some order in $1/N$. The computations are simplified if
dimension two auxiliary fields are first used to eliminate the quartic
couplings [22,23].

Since the scaling $(22i)$ only confirms our perturbative analysis we will just
state how the analysis proceeds. $V_{eff}$ at $O(N)$ depends only
on the combination $\phib\cdot\psi$ both at tree level and after all
the $O(N)$ diagrams are included. The real part of $V_{eff}$ is an
increasing function of $\phib\cdot\psi$ and therefore in the large $N$ limit
the vacuum energy is minimized at $\phib\cdot\psi=0$. In this limit,
$O(1)$ corrections are easy enough to compute; they include the cubic
term
$  - {T\over 12\pi} \[ {\alpha\over 3}(\phib^2+\psi^2) \]^{3\over 2} $
which leads to first order behaviour. Note that unlike the perturbative
result (21) the cubic term contains no $T^2-T_2^2$ term. This is because
such a term is lower order in $N$.

To study the scaling (22ii) we add the following terms involving
dimension two auxiliary fields,
$\chi_+$ and $\chi_-$, to the tree potential (23ii):
\be
     -{3N\over 2\lambda T}
    \({\chi_+ + \chi_-\over 2}-{\lambda T\over 6}(\vp^2+\vt^2) -m^2\)^2
      -{3N\over 2\lambda T}
    \({\chi_+ -\chi_-\over 2}-{\lambda T\over 3}\vp\cdot
    \vt\)^2
\ee
after which the potential becomes to $O(N)$
\be
 V = -{3N\over 4\lambda T} (\chi_+^2+\chi_-^2)+{N\over2}
       \vp\cdot\vt(\chi_+-\chi_-) +{3N\over2}
         (\chi_++\chi_-)\({1\over 6}(\vp^2+\vt^2)+ {M^2\over\lambda T}\),
\ee
where $M^2= T^2\lambda/36-\lambda v^2/6$.
After the shift (24), the Lagrangian is only quadratic in the quantum
scalar fields; these can then be integrated out at one--loop to give the
leading order effective potential in terms of $\vp,\vt$ and $\chi_\pm$.
Eliminating the auxiliary fields from this effective potential by their
equations of motion gives $V_{eff}(\vp,\vt)$ to $O(N)$.

The eigenvalues of the scalar field mass matrix of (26) are just
$\chi_+$ and $\chi_-$. Both eigenvalues have multiplicity $N$. Therefore,
the $O(N)$ quantum corrections are
\be {N\over2} {\rm Tr}\ln[-\vec{\partial}^2+\chi_+] +{N\over2}
    {\rm Tr}\ln[-\vec{\partial}^2+\chi_-]. \ee
The field dependent divergent part can be absorbed by renormalization
of the tree potential and the remaining
field dependent finite part yields $(\chi_\pm \geq 0)$
\be V_{eff}=V-{N\over 12\pi}\( \chi_+^{3\over2}+\chi_-^{3\over2}\).\ee
Here, $V$ is given by (26).
The equations of motion for $\chi_\pm$ are then
\be \chi_\pm= {\lambda\over 6}(\phib\pm\psi)^2+M^2
      -{\lambda T\over 12\pi} \sqrt{\chi_\pm}. \ee

It is now straightforward to verify several facts. For $M^2\geq 0$,
$\chi_\pm\geq 0$ and the minimum of the potential is at $<\phib>=
<\psi>=0$. For $M^2<0$, $\chi_\pm$ are both zero at $\phib\cdot\psi=0$
and $\lambda (\phib^2+\psi^2)/6+M^2=0$. Now since
$  {d V_{eff}\over d (\phib\pm\psi)^2} =
       {\partial  V_{eff}\over \partial (\phib\pm\psi)^2} \propto \chi_\pm,$
the point $\chi_+=\chi_-=0$ has the interpretation [22,23] as the
symmetry breaking minimum for $M^2<0$. This point moves continuously
away from the origin as $T^2$ falls continuously below $6v^2$; this is
the signal of a second order transition.

Although we have only demonstrated that (2) has a second order transition
for the particular case $\lambda=\xi=\alpha$ and at large $N$, we
might expect a similar outcome for $N=4$ and $\alpha\leq O(\lambda,\xi)$,
i.e. the first order behavior driven by $\alpha$ is wiped out by the
quantum fluctuations due to $\lambda,\xi$ for small enough $\alpha$.
A completely reliable determination may only be possible on the lattice.

3) Finally, in this section, we determine more precisely the
ratio $\alpha/\lambda$ for which perturbation theory breaks down completely
and the maximum ratio $\alpha^2\over\lambda$ for which the effective 3d
theory can be trusted. We examine the one--loop 3d potential for
$\phib\cdot \psi=0$,  $\lambda=\xi$ and dropping for the time
being the $T^2-T_2^2$ part of the  "cubic" term:
\be V_{eff}={(\lambda T)\over 4!}(\vp^2+\vt^2)^2+{M^2\over 2} (\vp^2+\vt^2)
  -{ \alpha^{3\over2}
      \over 12\pi} \[{1\over3}(\phib^2+\psi^2)
     \]^{3\over2}.\ee
$M^2$ and $T_2^2$ are given by (15) and (16), respectively.
For this potential,
$  T_1^2-T_2^2 = {\alpha^3 T_1^2 / 9\lambda\pi^2
     (\alpha+5\lambda)}, $ and
$  M(T_1) = {\sqrt{\alpha^3/\lambda} T_1 / 18\pi}$  .

Now note that the high $T$ perturbative
expansion parameters for (2) are only proportional to
$\lambda T/m,\xi T/m, \alpha T/m$. To get an estimate of
the constants of proportionality we examined higher loop corrections in
the 3d theory which grow with the number of scalars. For $\xi = \lambda$,
the large $N$ loop expansion parameters for $N=4$ are
$ {\lambda T/ ( 6\pi m)},{\alpha T/ ( 6\pi m)}, $
where $m^2$ is a field dependent mass eigenvalue.
At $T_1$ these expansion
parameters are $\approx 1$ for $\alpha\approx 5\lambda$. In addition,
at $T=T_1$ and for
$\alpha\geq 5\lambda$ the $T^2-T_2^2$ term is less than the $\phib^2+\psi^2$
term in the last term of (30), i.e. (30) gives a good description of
the potential.

The condition that all field dependent mass eigenvalues are less
than $2\pi T$ at the minimum at $T=T_2$ gives $\alpha^2 < 12\pi^2 \lambda$.
Thus, we predict first order behaviour for around\footnote{We
remark that as long as all couplings are perturbative at zero T
these conditions for first order behaviour are not unnatural.}
\be {\alpha^2\over 12\pi^2} < \lambda < {\alpha \over 5}. \ee
 For $\xi=0$ this last lower bound on $\alpha$
is instead around $\lambda <  \alpha/3$.

{\bf The Two Higgs Doublet Model}. As we have mentioned, the general
two Higgs doublet model has 5 scalar quartic couplings, not three. In addition,
it has  two scalar $vev$s $v_1,v_2$ , gauge couplings, and
dimension two terms which softly break
the discrete symmetry $\phib\rightarrow -\phib$.
The scalar potential is typically written [20]
\def\Phib{\Phi_1}\def\Psi{\Phi_2}
\begin{eqnarray}
 V&=& \lambda_1 (\Phib^{\dag}\Phib-v_1^2)^2+
    \lambda_2 (\Psi^{\dag}\Psi-v_2^2)^2  \nonumber \\
   & & + \lambda_3 \[ \Phib^{\dag}\Phib+\Psi^{\dag}\Psi-v_1^2
       -v_2^2 \]^2
     +\lambda_4 \[ \Phib^{\dag}\Phib\Psi^{\dag}\Psi
        -\Phib^{\dag}\Psi\Psi^{\dag}\Phib \]
       \nonumber \\
   & &+ \lambda_5 \[ {\rm Re}(\Phib^{\dag}\Psi)-v_1 v_2 \cos\zeta\]^2
      + \lambda_6 \[ {\rm Im}(\Phib^{\dag}\Psi)-v_1 v_2 \sin\zeta\]^2,
\end{eqnarray}
where the $\lambda_i,\zeta$ are real parameters and $\Phib,\Psi$ are
complex Higgs doublets. We have the following identifications
(when $\lambda_1=\lambda_2$, $\lambda_4=\lambda_6=0$)
\be \lambda_1+\lambda_3 =\lambda/6,\quad \lambda_3=\xi/6,\quad
    \lambda_5=2\alpha/3, \ee
between  (2) and (32) with the normalization $\Phib^{\dag}\Phib=\phib^2/2$,
$\Psi^{\dag}\Psi=\psi^2/2$.

It is reasonable to ask what can be learnt about the
the transition in the two doublet model from the analysis of
our toy model. One can convince oneself that
(32) will display first order behaviour with the following assumptions:
i) $v_1=v_2$, ii) $\lambda_5$ is big enough compared to the
other quartic scalar couplings and gauge couplings,
and iii) as with the one doublet model, there is
a nonperturbatively generated mass for the nonabelian magnetic gauge
fields [27]. Conditions i) and ii) ensure that the mechanism for first
order behaviour (within perturbation theory) is the same as for our
toy model and condition iii) is necessary to ensure that
the loop expansion parameter associated with the nonabelian gauge fields
is small at small scalar field values at $T=T_1$. Although our
toy model (2) has $\phib\cdot \psi=0$ at tree level, the fact that
the $O(\epsilon)$ RG results are independent of terms quadratic in
the fields in the tree potential strongly suggests (but does not prove)
that  we need not require $\cos\zeta$ to be
small for (32) to exhibit first order behaviour ($\cos\zeta\neq 1$
generically implies CP violation in the Higgs sector). The perturbative
analysis for $\cos\zeta\approx 1$ is more subtle than for our toy model.

 However,  note that for
$\sin\zeta\approx 0$, $\lambda_6({\rm Im}\Phib^{\dag}\Psi)^2=
{1\over4} \lambda_6 (\phib T \psi)^2$, with nonzero entries
$T_{1,2}=T_{3,4}=-T_{2,1}=-T_{4,3}=1$. We have checked that this
term can play the same role as the last term in (2). Thus, we predict
first order behaviour when $\sin\zeta\approx 0$ and $\lambda_6$
(which determines the pseudoscalar mass $M_A$) is sufficiently larger than
the other couplings, even when the mass of the lightest neutral Higgs
$M_h$ is near or above $M_W$. The precise
determination of the parameter space for which first order behaviour
is possible will require a more careful analysis, however the ratio
$M_A/M_h$ could be as low as 3.

{\bf Acknowledgements.} We are grateful to Rob Leigh and especially
Raman Sundrum for interesting and useful discussions.

{\bf Note Added.} We have just received an interesting
article on the electroweak phase transition with a singlet:
J.R. Espinosa and M. Quir\'os, IEM-FT-67/93, hep-ph/9301285.

\end{document}